\documentclass
[aps,pra,superscriptaddress,floatfix,showpacs,10pt,twocolumn]{revtex4}%
\usepackage{amsmath}
\usepackage{graphicx}
\usepackage{amsfonts}
\usepackage{amssymb}
\usepackage[T1]{fontenc}
\usepackage[latin2]{inputenc}
\usepackage{latexsym}
\usepackage{bbm}
\usepackage{color}%
\begin{document}
\title{Purifying entanglement of noisy two-qubit states via entanglement swapping}
\author{Wei Song}
\affiliation{School of Electronic and Information Engineering, Hefei Normal University,
Hefei 230061, China}
\author{Ming Yang}
\affiliation{School of Physics and Material Science, Anhui University, Hefei 230601, China }
\author{Zhuo-Liang Cao}
\affiliation{School of Electronic and Information Engineering, Hefei Normal University,
Hefei 230061, China}
\date{\today}

\begin{abstract}
Two qubits in pure entangled states going through separate
paths and interacting with their own individual environments will gradually
lose their entanglement. Here we show that the entanglement change
of a two-qubit state due to amplitude damping noises can be
recovered by entanglement swapping. Some initial states can be asymptotically
purified into maximally entangled states by iteratively using our protocol.

\end{abstract}
\maketitle

Quantum entanglement has received a lot of attentions due to its
importance in almost all quantum informational processing and
communication tasks \cite{Neilsen:2000}. To actually realize such
tasks, highly entangled states are usually required. However, the
quality of entanglement will degrade exponentially due to the
unavoidable interactions between a quantum system and its
environment. Entanglement purification provides a way to extract a
small number of entangled pairs with a relatively high degree of
entanglement from a large number of less entangled pairs using only
local operations and classical communication (LOCC). If the less
entangled states are pure, we call such a process entanglement
concentration, while for mixed states case, this process is usually
called entanglement purification or distillation. The first
entanglement purification protocol for mixed states was introduced
by Bennett \emph{et al} in 1996 \cite{Bennett:1996}. Since then,
many theoretical and experimental works have been made in the field
of entanglement purification and
concentration\cite{Deutsch:1996,Horodecki:1997,Linden:1998,Kent:1998,Duan:2000,Pan:2001,Dur:2003,Nakazato:2003,Bose:1999,Zhao:2001,Yamamoto:2001,Pan:2003,Fan:2009,Yang1:2005}%
. Especially Bose \emph{et al.}\cite{Bose:1999} have proposed a
scheme for concentration of pure shared entanglement via
entanglement swapping \cite{Zukowski:1993}. The idea is that Alice
shares an entangled state with Bob and Bob shares the same entangled
state with Charlie. Bob performs a Bell measurement on his two
particles, and then Alice and Charlie end up sharing a final state
with a greater entanglement than that of the initial states
probabilistically. However, their discussion only works for the pure
state cases where both initial states are of the form $\alpha\left|
{00} \right\rangle + \beta\left|  {11} \right\rangle $. It is well
known that entanglement swapping is one of the central ingredients
for quantum repeaters, which lays at the heart of quantum
communication \cite{Briegel:1998,Modlawska:2008,Khalique:2013}. If
we can extend the idea of entanglement swapping to the purification
of mixed state cases, then Bell state measurements can be
substituted for the original CNOT operations in the purification
process. At present, for the general unknown mixed states, whether
entanglement swapping can be used to purify the entanglement remains
unknown. The main difficulty lies in the different structures of
state spaces for pure states and mixed states, and thus the
discussion of pure state cases cannot be generalized to the mixed
cases directly. In this paper we will explore the problem by
considering some concrete examples. We find that the entanglement
degradation of a two-qubit state due to amplitude damping(AD) noises
can be recovered by entanglement swapping combined with weak
measurements. Here, we take the AD noises to model the environment,
and such noise is one important type of decoherence which is related
to many practical systems \cite{Apollaro:2010}. It has a useful
physical interpretation: there is some probability of decaying from
state $\left| 1 \right\rangle$ to $\left| 0 \right\rangle$, but it
never transforms state $\left| 0 \right\rangle$ to $\left| 1
\right\rangle$. For example, photon loss in an optical fiber can be
described by this model\cite{Shor:2011}. Recently, several
interesting methods have been put forward to cope with this type of
noise using weak measurements followed by quantum measurement
reversals
 \cite{Kim:2012,Sun:2010,Man:2012}or error-correcting codes
\cite{Shor:2011,Fletcher:2008}. Compared with these works, our
method provides a different way for battling against decoherence
from AD noises. The distinct advantage of our scheme is that the
final state is almost maximally entangled state after several rounds
of our protocol, while the protocol in Ref.
\cite{Kim:2012,Sun:2010,Man:2012} is a one-round scheme but also depends on
the initial states to be protected. Furthermore, we do not require
the singlet fraction larger than $ \frac{1}{2}$ in contrast to the
BBPSSW protocol\cite{Bennett:1996}.

We now explain our scheme by discussing a specific example firstly.
Suppose two pairs of qubits are initially prepared in the states
$|\phi\rangle=\sqrt a \left|  {01} \right\rangle + \sqrt{1 - a}
\left|  {10} \right\rangle $ and $|\tilde{\phi}\rangle=\sqrt a
\left| {10} \right\rangle + \sqrt{1 - a} \left|  {01} \right\rangle
$, respectively. Note that these two pairs have the same
entanglement and the second pair is just the flipped state of the
first one. One pair is transmitted through the local individual AD
channels to two separated user Alice and Bob and the other pair is
transmitted through the local individual AD channels to Bob and
Charlie. For simplicity, we suppose the local AD channels are all
identical in the following arguments. The initial pure states will
evolve into mixed states under the disturbance of AD noises. This
process can be described by a completely positive trace preserving
map $ \rho\left(  t \right)  = \varepsilon\left( {\rho\left(  0
\right) } \right)  = \sum\nolimits_{\mu }{K_{\mu}\left(  t \right)
\rho\left( 0 \right) } K_{\mu}^{\dag}\left(  t \right) $, where the
operators $ \left\{ {K_{\mu}\left(  t \right) } \right\} $ satisfy
the completeness condition $\sum\nolimits_{\mu}{K_{\mu}^{\dag}\left(
t \right) K_{\mu}\left(  t \right) } = \mathbb{I}$. In the case of
AD noises the Kraus operators $K_{1} = \left|  0 \right\rangle
\left\langle 0 \right|  + \sqrt{1 - p} \left|  1 \right\rangle
\left\langle 1 \right| , K_{2} = \sqrt p \left|  0 \right\rangle
\left\langle 1 \right| $. After the AD nosies, Alice and Bob will
share the mixed state $\rho_{AB} =  p|00\rangle\langle
00|+(1-p)|\phi\rangle\langle\phi|$, and Bob and Charlie will share
the state $\rho_{BC} =  p|00\rangle\langle
00|+(1-p)|\tilde{\phi}\rangle\langle\tilde{\phi}|$. Now Bob performs
a Bell state measurement on his qubits from the two mixed pairs in
the basis $\left\{  {\left| {\Phi^{\pm}} \right\rangle =
\frac{1}{\sqrt2 }\left(  {\left|  {00} \right\rangle \pm\left|  {11}
\right\rangle } \right) ,\left| {\Psi^{\pm}} \right\rangle =
\frac{1}{\sqrt2 }\left(  {\left|  {01} \right\rangle \pm\left|  {10}
\right\rangle } \right) } \right\} $. In order to realize a
purification process, the entanglement of the final state after
entanglement swapping should be larger than that of the initial
states. Here, we use the Wootter's concurrence \cite{Wootters:1998}
to quantify the entanglement: $C\left( \rho\right)  = \max\left\{
{0,\sqrt{\lambda_{1} } - \sqrt{\lambda_{2} } - \sqrt{\lambda_{3} } -
\sqrt{\lambda_{4} } } \right\} $, where ${\lambda_{i} }$ are the
eigenvalues in decreasing order of the matrix $ \rho\tilde\rho=
\rho\sigma_{y} \otimes\sigma_{y} \rho^{*} \sigma_{y}
\otimes\sigma_{y}$ with $\rho^{*}$ denoting the complex conjugation
of $\rho$. After simple calculation, we can see that the two mixed
states have the same concurrence $ C\left( {\rho_{AB} } \right)  =
C\left( {\rho_{BC} } \right)  = 2\left(  {1 - p} \right)
\sqrt{a\left(  {1 - a} \right) }$. If Bob obtains $\left|
{\Psi^{\pm }} \right\rangle $ as the result of his Bell state
measurement, the state of Alice and Charlie will collapse into the
mixed state:
\begin{align}
\rho_{AC} = \frac{1}{N}\left(  {2p\left(  {1 - p} \right) a\left|  {00}
\right\rangle } \right. \left\langle {00} \right| \nonumber\\
\left.  { + 2\left(  {1 - p} \right) ^{2} a\left(  {1 - a} \right) \left|
{\Psi^{\pm}} \right\rangle \left\langle {\Psi^{\pm}} \right| } \right)
\end{align}

\begin{figure}[ptb]
\includegraphics[scale=0.55,angle=0]{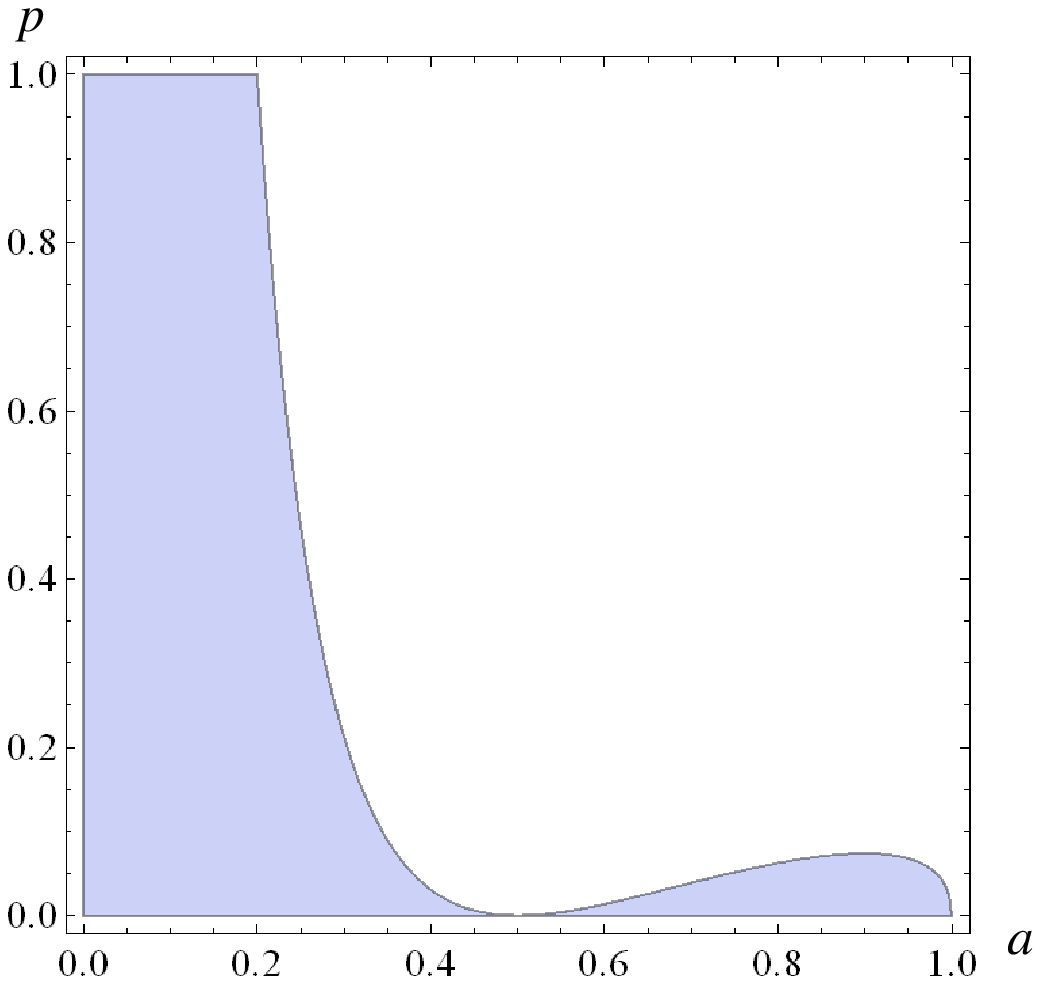}\caption{(color online). If the
entanglement of the final state is larger than the initial state, then the
parameters $p$ and $a$ should lie in the blue region.}%
\end{figure}

\noindent where $N = 2\left(  {1 - p} \right) ^{2} a\left(  {1 - a}
\right)  + 2p\left(  {1 - p} \right) a$. The probabilities for obtaining
$ \left|  {\Psi^{+} } \right\rangle $ or $\left|  {\Psi^{-} }
\right\rangle $ are $\frac{1}{2}N $, respectively. The concurrence
of this state is $\frac{2}{N}\left(  {1 - p} \right) ^{2} a\left( {1
- a} \right)  $. Then an enhancement of entanglement implies that $
C\left(  {\rho_{AC} } \right)  > C\left(  {\rho_{AB} } \right) $. We plot the region of the parameters $p$ and $a$ satisfying
the inequality  $ C\left(  {\rho_{AC} } \right)  > C\left(
{\rho_{AB} } \right) $ in Fig.1. We find the region is not empty, i.e.,
if $p$ and $a$ are restricted in the blue region, the entanglement
of the final mixed state can be larger than that of the original states. If $0 \le a \le0.2$ the mixed state entanglement from arbitrary AD noises always
can be enhanced through the above entanglement
swapping process. If Bob obtains $\left| {\Phi^{\pm}} \right\rangle
$ as the results of his measurements, then entanglement of the final
state cannot be larger than that of the initial one. It is worth pointing out that, if we do
not flip the second pair at the beginning, then Alice and Charlie
will share a mixed state with entanglement less than that of the original
state. In real situations, the initial two pairs are not in the same
form usually. Suppose that the second pair is in the form
$\sqrt{a^{\prime}} \left|  {10} \right\rangle + \sqrt{1 -
a^{\prime}} \left|  {01} \right\rangle $ which has a small deviation
from the first pair, where $a^{\prime}\ne a $. Similar calculations
show that if Bob obtains $\left|  {\Psi^{\pm }} \right\rangle $ as
the result of his Bell state measurement, then Alice and Charlie will share the
mixed state:

\begin{eqnarray}
&\frac{1}{M}\left( {p\left( {1 - p} \right)} \right.\left( {a + a'}
\right)\left| {00} \right\rangle \left\langle {00} \right| + \left(
{1 - p} \right)^2 a\left( {1 - a'} \right)\left| {01} \right\rangle
\left\langle {01} \right|&\nonumber\\
& + \left( {1 - p} \right)^2 a'\left( {1 - a} \right)\left| {10}
\right\rangle \left\langle {10} \right| \pm \left( {1 - p} \right)^2
 \times&\nonumber\\
&\sqrt {aa'\left( {1 - a} \right)\left( {1 - a'} \right)} \left(
{\left| {01} \right\rangle \left\langle {10} \right| + \left| {10}
\right\rangle \left\langle {01} \right|} \right)&
\end{eqnarray}

\noindent where $M = p\left(  {1 - p} \right) \left(  {a +
a^{\prime}} \right) + \left(  {1 - p} \right) ^{2} \times\\\left(
{a^{\prime}\left(  {1 - a} \right)  + a\left(  {1 - a^{\prime}}
\right) } \right)$. We plot the possible region of the states whose
entanglement can be enhanced via the entanglement
swapping process. From Fig.2 one can see that our scheme is still
feasible if the parameters $a^{\prime}$ and $a$ are restricted in
the blue region. It shows that small deviations of the initial pairs
are allowable in above swapping process.

\begin{figure}[ptb]
\includegraphics[scale=0.72,angle=0]{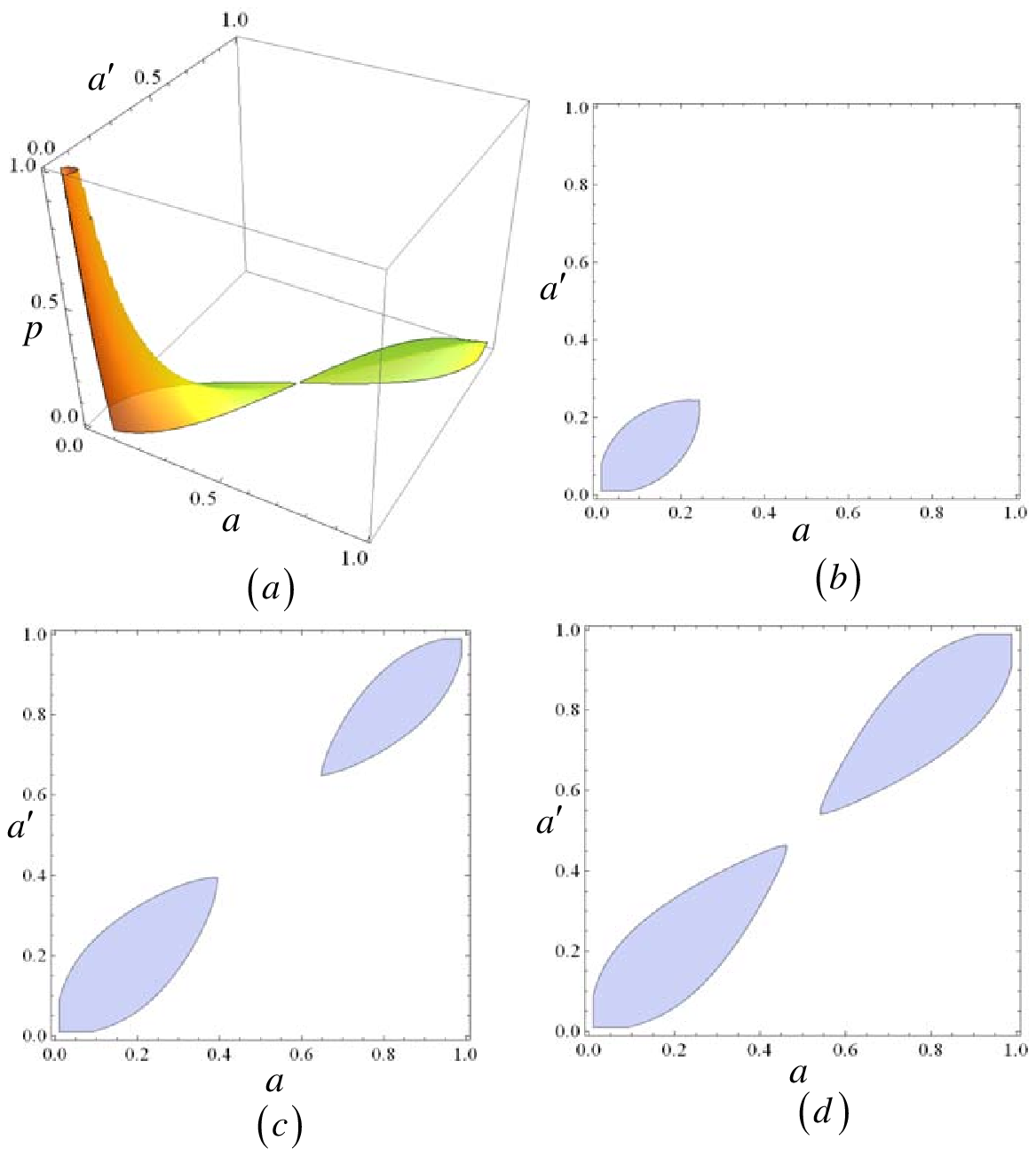}\caption{(color online). The
region of the states whose entanglement can be enhanced through
entanglement swapping with different parameters, respectively. (b)
$p = 0.1$,(c) $p =
0.01$,(d) $p = 0.001$.}%
\end{figure}

Next we will consider whether we can enlarge the range of
the possible parameters $p$ and $a$ in Fig.1. In order to further
enhance the entanglement of the initial state, the first choice is to
apply the procedure described above several times. However, if we start the second round of the protocol with the mixed state $\rho_{AC}$, one can directly get that its entanglement will
degrade after the protocol. Suppose two copies of the state
$\rho_{AC}$ in Eq.(1) have been prepared through the entanglement swapping
process. Before the second round of our protocol, the
weak nondestructive measurements $M_{\pm}$ will be performed on one of the qubits of
$\rho_{AC}$, where $M_{+} = \sqrt b \left|  0 \right\rangle
\left\langle 0 \right|  + \sqrt{1 - b} \left|  1 \right\rangle
\left\langle 1 \right| ,M_{-} = \sqrt{1 - b} \left|  0 \right\rangle
\left\langle 0 \right|  + \sqrt b \left|  1 \right\rangle
\left\langle 1 \right| $. Obviously, this operation increases the fraction of $\left| 1 \right\rangle$ or $\left| 0
\right\rangle$. If the weak measurement outcomes on the qubit $A$ of
the first pair is $M_{+}$, the mixed state will be tranformed to the following one:

\begin{eqnarray}
&&\rho '_{AC}  =\nonumber\\
&&\frac{1}{{N'}}\left( {2p\left( {1 - p} \right)ab\left| {00}
\right\rangle \left\langle {00} \right| + \left( {1 - p} \right)^2
a\left( {1 - a} \right)\left| \varphi\right\rangle \left\langle
\varphi  \right|} \right)\nonumber\\
\end{eqnarray}

\noindent where $N^{\prime}= 2p\left(  {1 - p} \right) ab + \left(
{1 - p} \right) ^{2} a\left(  {1 - a} \right) $, $ \left| \varphi
\right\rangle  = \sqrt b \left| {01} \right\rangle  + \sqrt {1 - b}
\left| {10} \right\rangle $. The probability for obtaining this state is
\begin{align*}
p^{\prime}= \frac{{2p\left(  {1 - p} \right) ab + \left(  {1 - p}
\right) ^{2} a\left(  {1 - a} \right) }}{{2p\left(  {1 - p} \right)
a + 2\left(  {1 - p}
\right) ^{2} a\left(  {1 - a} \right) }}%
\end{align*}

If the weak measurement on the qubit C of the second pair is also
$M_{+}$, the mixed state will be transformed to the following one:

\begin{eqnarray}
&&\rho ''_{AC}  =\nonumber\\
&&\frac{1}{{N''}}\left( {2p\left( {1 - p} \right)ab\left| {00}
\right\rangle \left\langle {00} \right| + \left( {1 - p} \right)^2
a\left( {1 - a} \right)\left| \tilde \varphi\right\rangle
\left\langle
\tilde \varphi  \right|} \right)\nonumber\\
\end{eqnarray}

\noindent where $ \left| \tilde\varphi \right\rangle  = \sqrt b
\left| {10} \right\rangle  + \sqrt {1 - b} \left| {01} \right\rangle
$, $N^{\prime\prime }=N^{\prime}$ and the corresponding probability
is $p^{\prime}$. Now we start the second round of purification process with these two new mixed states. Bob
performs the Bell state measurements on the two qubits, and the
resulting state is $\rho_{AC}^{\left(  2 \right) } = \frac
{1}{{N^{\left(  2 \right) } }}\left(  {4p\left( {1 - p} \right) ^{3}
b^{2} a^{2} \left(  {1 - a} \right) \left| {00} \right\rangle
\left\langle {00} \right| } \right.   + 2b\left( {1 - b} \right)
\left(  {1 - p} \right) ^{4} \left.  {a^{2} \left( {1 - a} \right)
^{2} \left|  {\Psi^{\pm} } \right\rangle \left\langle {\Psi^{\pm} }
\right| } \right)  $ if the measurement result is ${\left|
{\Psi^{\pm} } \right\rangle }$, where the superscript $(2)$ of $\rho^{(2)}_{AC}$
denotes the second round of our protocol. The concurrence of ${\rho
_{AC}^{\left(  2 \right) } } $ is
\begin{widetext}
\begin{eqnarray}
C\left( {\rho _{AC}^{\left( 2 \right)} } \right) = \frac{2{b\left(
{1 - b} \right)\left( {1 - p} \right)^4 a^2 \left( {1 - a} \right)^2
}}{{4p\left( {1 - p} \right)^3 b^2 a^2 \left( {1 - a} \right) +
2b\left( {1 - b} \right)\left( {1 - p} \right)^4 a^2 \left( {1 - a}
\right)^2 }}
\end{eqnarray}
\end{widetext}

\begin{figure}[ptb]
\includegraphics[scale=0.74,angle=0]{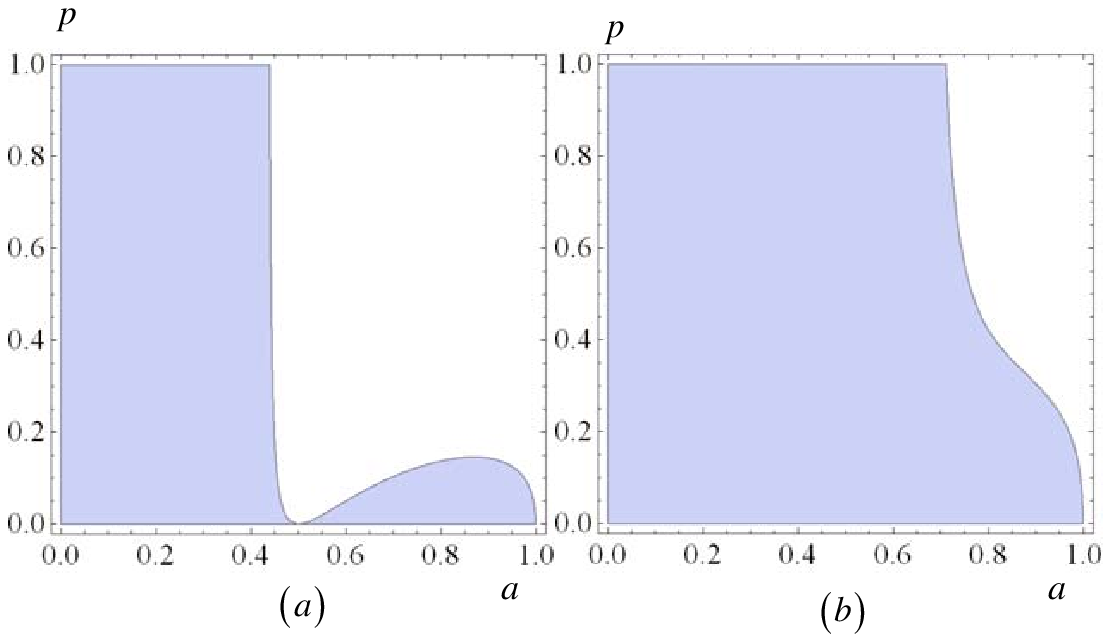}\caption{(color online). The
region of the states whose entanglement can be enhanced through
entanglement swapping with different $n$, respectively. (a) $n = 2$,(b) $n = 3$.}%
\end{figure}

The probability for obtaining this state is $p\left(
{\rho_{AC}^{\left( 2 \right) } } \right)  = 2p\left(  {1 - p}
\right) ^{3} b^{2} a^{2} \left(  {1 - a} \right)  + b\left(  {1 - b}
\right) \left(  {1 - p} \right) ^{4} a^{2} \left(  {1 - a} \right)
^{2} $. For fixed $p$ and $a$, the concurrence is close to 1 if $b$
close to $0$, and the probability is close to 0 too. There exists a
tradeoff between the probability and the concurrence: $C\left(
{\rho_{AC}^{\left(  2 \right) } } \right)  \cdot p\left(
{\rho_{AC}^{\left(  2 \right) } } \right)  = b\left(  {1 - b}
\right) \left(  {1 - p} \right) ^{4} a^{2} \left(  {1 - a} \right)
^{2}$. The entanglement of the final state can be increased further
as long as the following relation holds: $C\left( {\rho_{AC}^{\left(
2 \right) } } \right)  > C\left( {\rho_{AC}^{\left(  1 \right) } }
\right)> C\left( {\rho_{AB} } \right) $. It is straightforward to
verify that $C\left( {\rho_{AC}^{\left(  2 \right) } } \right)  >
C\left( {\rho_{AC}^{\left(  1 \right) } } \right) $ corresponds to
the inequality $ \frac{{\left( {a - 1} \right)\left( {p - 1}
\right)\left( {1 - 3b} \right)}}{{\left( {1 + a\left( {p - 1}
\right)} \right)\left( {\left( {1 - b} \right)\left( {1 - p}
\right)\left( {1 - a} \right) + 2bp} \right)}} > 0 $, and this
inequality holds when $  b < \frac{1}{3}$. We have to mention that
we cannot choose the condition $b=0$ because the weak measurement
becomes a projective measurement and the resulting states
$\rho^{\prime}_{AC}$ and $\rho^{\prime\prime}_{AC} $ are both
separable states in this case.

By exchanging $b$ and $1-b$ in Eq.(5) we can get the concurrence of
${\rho _{AC}^{\left( 2 \right)} }$ if the weak measurements on the
qubits $A$ and $C$ of the two pairs are both $M_{-}$. In this case,
the inequality $C\left( {\rho_{AC}^{\left(  2 \right) } } \right)  >
C\left( {\rho_{AC}^{\left(  1 \right) } } \right)$ implies $ b >
\frac{2}{3}$. If the weak measurements on the qubits $A$ and $C$ of
the two pairs are different, we find that the entanglement of the
final state cannot be larger than that of the initial one. Thus we
have shown that entanglement can be further purified in the second
round of our protocol. If the measurement results of Bob are $\left|
{\Phi^{\pm}} \right\rangle $, then the concurrence of ${\rho
_{AC}^{\left( 2 \right)} }$ cannot be further increased.

By iteratively using this procedure several times, the final mixed
state becomes $\rho_{AC}^{\left( n \right) } = \frac{1}{{N^{\left( n
\right) } }}\left(  {a_{n} \left| {00} \right\rangle \left\langle
{00} \right| + 2b_{n} \left| {\Psi^{\pm} } \right\rangle
\left\langle {\Psi^{\pm} } \right| } \right) $, with $ a_{n} = 2^{n}
pb^{2^{n - 1} + n - 2} \left(  {1 - b} \right) ^{2^{n - 1} - n}
\left( {1 - p} \right) ^{2^{n} - 1} a^{2^{n - 1} } \left( {1 - a}
\right) ^{2^{n - 1} - 1}$, and $b_{n} = b^{2^{n-1} - 1} \left( {1 -
b} \right) ^{2^{n-1} - 1} \left(  {1 - p} \right) ^{2^{n} }
a^{2^{n-1} } \left( {1 - a} \right) ^{2^{n-1} }$ after $n$th round
of our protocol, where we have supposed the measurement results of
Bob are $\left| {\Psi^{\pm} } \right\rangle $ and the weak
measurements on the qubits $A$ and $C$ of the two pairs are both
$M_{+}$. The entanglement of the final state can be larger than the
initial one if the following series of inequalities hold: $C\left(
{\rho_{AC}^{\left( n \right) } } \right)
> \cdots>C\left( {\rho_{AC}^{\left(  2 \right) } } \right)  >
C\left( {\rho_{AC}^{\left(  1 \right) } } \right)  > C\left(
{\rho_{AB} } \right) $. Numerical calculation shows that a
near-perfect maximally entangled state can be extracted in the limit
of infinite rounds for arbitrary $p$ and $a$. We plot the possible
region for the cases $n=2$ and $n=3$ with fixed parameter $b=0.22$,
respectively in Fig.3. Obviously, the possible region of parameters
$p$ and $a$ becomes larger with the increase of $n$. In Fig.4 we
also plot  the concurrence of the states $\rho _{AB} $, $\rho
_{AC}^{\left( 1 \right)} $, $\rho _{AC}^{\left( 2 \right)} $ and
$\rho _{AC}^{\left( 3 \right)}$, respectively for $a=0.3, b=0.22$.

\begin{figure}[ptb]
\includegraphics[scale=0.8,angle=0]{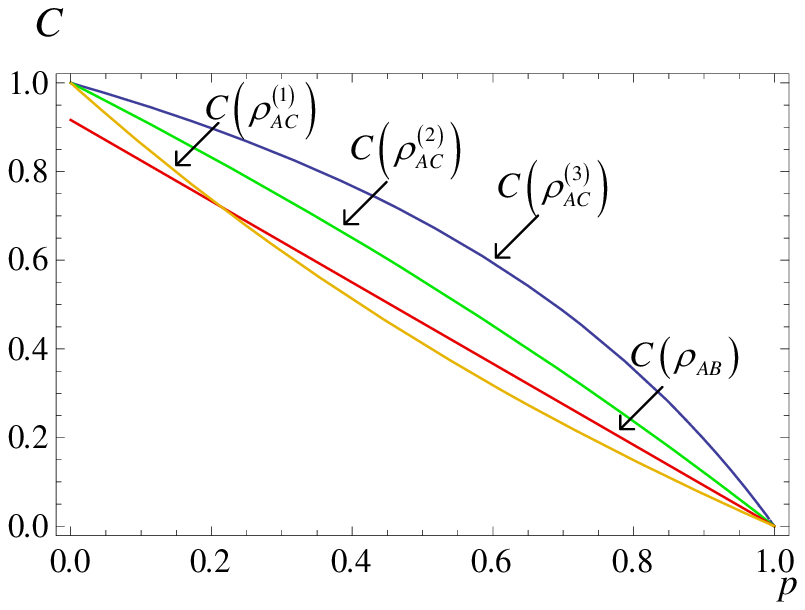}\caption{(color online). Plots of the
concurrence of the states $\rho _{AB} $ (red line) , $\rho _{AC}^{\left( 1 \right)} $ (yellow line), $\rho _{AC}^{\left( 2 \right)} $ (green line) and $\rho _{AC}^{\left( 3 \right)}$ (blue line) for $a=0.3, b=0.22$.}%
\end{figure}

In the following we consider the case when the two states
initially prepared in the other form $\sqrt A \left|  {00}
\right\rangle + \sqrt{1 - A} \left|  {11} \right\rangle $. One pair
is distributed to two separated users Alice and Bob and the other
pair to Bob and Charlie via the amplitude damping(AD) noisy channels. The states evolve into $\chi_{AB} = \chi_{BC} = \left[  {A +
\left(  {1 - A} \right) p^{2} } \right] \left|  {00} \right\rangle
\left\langle {00} \right|  + \left(  {1 - A} \right) p\left(  {1 -
p} \right) \left(  {\left|  {01} \right\rangle \left\langle {01}
\right|  + \left|  {10} \right\rangle \left\langle {10} \right| }
\right)   + \left(  {1 - A} \right) \left(  {1 - p} \right) ^{2}
\newline\times\left|  {11} \right\rangle \left\langle {11} \right|
+ \left(  {1 - p} \right) \sqrt{A\left(  {1 - A} \right) } \left(
{\left|  {00} \right\rangle \left\langle {11} \right|  + \left| {11}
\right\rangle \left\langle {00} \right| } \right)  $. Let Bob
performs the Bell measurements on his qubits. If Bob obtains $\left|
{\Psi^{\pm} } \right\rangle $ as the result of his measurement,
the shared state between Alice and Charlie is%

\begin{align}
\chi_{AC}  & =\left(  {1-A}\right)  \left(  {1-p}\right)  ^{2}\left(
{A+2\left(  {1-A}\right)  p^{2}}\right)  \nonumber\\
& \times\left(  {\left\vert {01}\right\rangle \left\langle
{01}\right\vert
+\left\vert {10}\right\rangle \left\langle {10}\right\vert }\right)  \nonumber\\
& +2 \left(  {1-A}\right)  p\left(  {1-p}\right)  \left( {A+\left(
{1-A}\right)  p^{2}}\right)  \left\vert {00}\right\rangle
\left\langle
{00}\right\vert \nonumber\\
& +2\left(  {1-A}\right)  ^{2}p\left(  {1-p}\right)  ^{3}\left\vert
{11}\right\rangle \left\langle {11}\right\vert \nonumber\\
& \pm\left(  {1-A}\right)  A\left(  {1-p}\right)  ^{2}\left(
{\left\vert {01}\right\rangle \left\langle {10}\right\vert
+\left\vert {10}\right\rangle \left\langle {01}\right\vert }\right),
\end{align}

\noindent where we have omitted the normalization factor. The
probability for obtaining this state is $p = \left(  {1 - A} \right)
\left(  {1 - p} \right) ^{2} \left(  {A + 2\left(  {1 - A} \right)
p^{2} } \right)  + \left(  {1 - A} \right) A\left(  {1 - p} \right)
^{2}$. In Fig.5(a) we plot the region for $C\left(  {\chi_{AC} }
\right)  > C\left(  {\chi_{AB} } \right) $. Similarly, if Bob
obtains $\left| {\Phi^{\pm}} \right\rangle $ as the result of his
measurement, then entanglement cannot be increased. In order to
further enlarge the region of the possible parameters in Fig.5(a),
Alice and Charlie will perform weak nondestructive measurements
$M_{\pm}$ on their qubits of $\chi_{AC}$, respectively. If the
measurement results are both $M_{+}$, the mixed state is transformed
to

\begin{align}
\chi'_{AC}  & =\left(  {1-A}\right)  \left(  {1-p}\right) ^{2}\left(
{A+2\left(  {1-A}\right)  p^{2}}\right)  \nonumber\\
& \times{b(1-b)}\left(  {\left\vert {01}\right\rangle \left\langle
{01}\right\vert
+\left\vert {10}\right\rangle \left\langle {10}\right\vert }\right)  \nonumber\\
& +2{b}^{2}\left(  {1-A}\right)  p\left(  {1-p}\right)  \left(
{A+\left( {1-A}\right)  p^{2}}\right)  \left\vert {00}\right\rangle
\left\langle
{00}\right\vert \nonumber\\
& +2{(1-b)}^{2}\left(  {1-A}\right)  ^{2}p\left(  {1-p}\right)
^{3}\left\vert
{11}\right\rangle \left\langle {11}\right\vert \nonumber\\
& \pm\sqrt {b\left( {1 - b} \right)}\left(  {1-A}\right)  A\left(
{1-p}\right)  ^{2}\left( {\left\vert {01}\right\rangle \left\langle
{10}\right\vert +\left\vert {10}\right\rangle \left\langle
{01}\right\vert }\right).
\end{align}

For $ 0 < b < \frac{1}{2}$, numerical calculations show that the
region can be further enlarged. We plot the possible region in
Fig.5(b) with $ b=0.25 $. If the measurement results are both
$M_{-}$, the parameter $ b $ must satisfy $ \frac{1}{2} < b <1 $.
While for cases with different weak measurement results, the region
cannot be enlarged. In order to have a vivid picture, we plot the
concurrence of the state $\chi_{AB} $, $\chi_{AC} $ and $\chi'_{AC}
$ in Fig.6, respectively, where we have chosen $A=0.9, b=0.25$. From
Fig.6 we can see that the concurrence of $\chi'_{AC} $ will be
always larger than that of the original mixed state $\chi_{AB} $.

\begin{figure}[ptb]
\includegraphics[scale=0.7,angle=0]{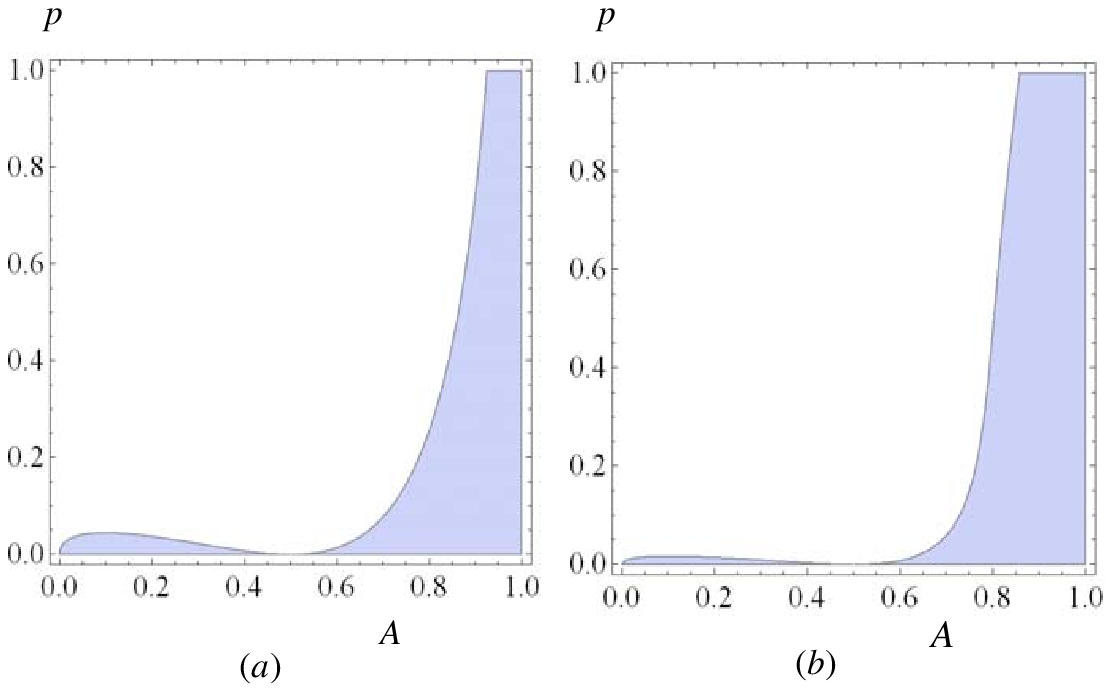}\caption{(color online). The
region of the states whose entanglement can be enhanced through
entanglement swapping if the initial state is $\chi_{AB}$. (a) The
final state is $\chi_{AC} $. (b) The final state is $\chi'_{AC} $.}%
\end{figure}

So far we have shown that entanglement swapping combined with weak
measurement can be used to purify a class of AD
noises induced mixed states. By a similar procedure, our protocol can be generalized
directly to the general two-qubit mixed states lying in the subspace
spanned by the basis $\left\{ {\left| {00} \right\rangle ,\left|
{01} \right\rangle ,\left| {10} \right\rangle } \right\}  $ or
$\left\{ {\left| {01} \right\rangle ,\left|  {10} \right\rangle
,\left|  {11} \right\rangle } \right\} $, if the conditions
$\rho_{22} \rho_{33} = \rho_{23} \rho_{32}$ is satisfied. An open
problem is whether entanglement swapping can be used to purify
arbitrary mixed states. If the answer is no, then can we find a
criteria to distinguish which states cannot be purified with
entanglement swapping. In practice, the entanglement swapping has
been demonstrated experimentally \cite{Pan:1998,Goebel:2008} and the
AD noise can be simulated by beam splitters with appropriate
transmission coefficients. Recently, the weak non-destructive
measurements have also been developed and demonstrated in
single-photon quantum optical systems\cite{Pryde:2004,Ralph:2006}.
Thus our protocol can be demonstrated experimentally within current
technology.

\begin{figure}[ptb]
\includegraphics[scale=0.8,angle=0]{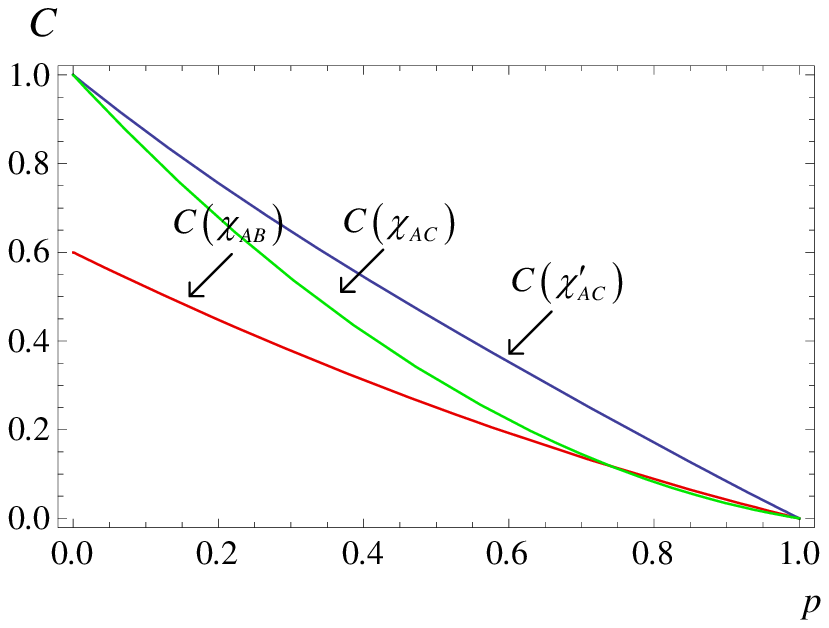}\caption{(color online). Plots of the
concurrence of the states $\chi_{AB} $ (red line) , $\chi_{AC} $ (green line) and $\chi'_{AC} $ (blue line) for $A=0.9, b=0.25$.}%
\end{figure}

This work was supported by National Natural Science Foundation of
China under Grant No.10905024, No.61073048, No.11274010,
No.11374085, the Key Project of Chinese Ministry of Education under
Grant No.211080, No.210092, the Specialized Research Fund for the
Doctoral Program of Higher Education(no. 20113401110002), the `211'
Project of Anhui University, the personnel department of Anhui
province and the Key Program of the Education Department of Anhui
Province under Grant No.KJ2011A243.

\end{document}